# Synthesis of Tungsten Oxide Nanoparticles using a Hydrothermal Method at Ambient Pressure

*Majid Ahmadi[1], Reza Younesi[2] and Maxime J-F Guinel[1, 3†]*

[1]Department of Physics, College of Natural Sciences, University of Puerto Rico, PO Box 70377, San Juan, PR 00936-8377, U.S.A.
[2]Department of Energy Conversion and Storage, Technical University of Denmark, Frederiksborgvej 399, P.O. Box 49, DK-4000 Roskilde, Denmark.
[3]Department of Chemistry, College of Natural Sciences, University of Puerto Rico, PO Box 70377, San Juan, PR 00936-8377, U.S.A.

**Abstract**
Tungstite ($WO_3 \cdot H_2O$) nanoparticles were synthesized using a simple and inexpensive low temperature and low pressure hydrothermal method by adding hydrochloric acid to diluted sodium tungstate solutions ($Na_2WO_4 \cdot 2H_2O$) at temperatures below 5°C. A heat treatment at temperatures at or above 300°C resulted in a phase transformation to monoclinic $WO_3$, while preserving the nanoparticles morphology. The products were characterized using powder x-ray diffraction, transmission electron microscopy (including electron energy-loss spectroscopy and electron diffraction) and x-ray photoelectron spectroscopy.

[†]*Corresponding author:* maxime.guinel@upr.edu

**Introduction**

Tungsten oxide ($WO_3$) is a transition metal oxide semiconductor with a widely tunable band gap, in the range of $E_g$=2.5-2.8eV at room temperature. Interest was recently put on $WO_3$ thin films and nanoparticles [1] for a wide variety of applications in microelectronics and optoelectronics [2], dye-sensitized solar cells [3], colloidal quantum dot LEDs [4], photocatalysis [5] and photoelectrocatalysis [6], water splitting photocatalyst as main catalyst [7-13] and methanol oxidation catalyst [14]. Environmental applications may also benefit with the use of $WO_3$ as a visible light photocatalyst to generate OH radicals in the wastewater treatment [15], bacteria destruction [16] and photocatalytic reduction of $CO_2$ into hydrocarbon fuels [17]. Yin *et al*. [18] have reported high hydrophobic properties and improved performances of $WO_3$ as anode materials in lithium ion batteries (LIB). $WO_3$ has also been used in so-called smart windows [19] for energy-efficient buildings, flat-panel displays, optical memory and writing-reading-erasing devices. Moreover, $WO_3$ shows excellent functional activity to various gases, such as $H_2$ [20], $H_2S$, $NO_x$, trimethylamine, and other organic compound gases such as acetone sensing in exhaled breath [21].

There exist few methods for the preparation of $WO_3$. For example using electrochemical processes [22], solution-based colloidal approaches [23], bioligation processes [24-25], and chemical vapor deposition techniques [26-27]. Oriented $WO_3$ nanowires were also obtained using an hydrothermal (HT) treatment in the presence of alkaline metal sulfates [28-29].

In this article, we report on a very simple, low temperature (95-98$^o$C) and low pressure (1bar) HT process to synthesize nanoparticles of $WO_3$. Therefore, the needs for a high pressure autoclave and highly acid-resistive liners in the reaction chamber are avoided, which present clear advantages for the industry. Moreover, it is possible to readily scale up this process for the production of large quantities of materials. A first reaction produces orthorhombic tungstite ($WO_3.H_2O$) followed by an HT process using a simple reflux system in the presence of oxalic acid as surfactant. Annealing at or above 300$^o$C in ambient air leads to a phase transformation to monoclinic $WO_3$, where the morphology of the materials is preserved.

**Experimental Details**

The tungstite ($WO_3.H_2O$) materials were synthesized using the acid precipitation method [30-31] followed by a low pressure (P=1bar) and low temperature (T=95-98$^o$C) HT treatment [32]. Ultrapure 18MΩ Millipore® deionized water was used for the preparation of all solutions. Around 30-50mL of 6N hydrochloric acid (HCl) were added drop-wise to a 100mL 15mM sodium tungstate solution ($Na_2WO_4.2H_2O$) while the solution was kept at below 5$^o$C and under constant stirring to produce white amorphous precipitates [33]. The tungstite nanoparticles (TNPs) were obtained by applying a HT treatment for 6-18 hours to that previous solution and by adding oxalic acid ($H_2C_2O_4$) as surfactant. This solution, yellowish in color, was centrifuged and washed several times to reach pH~6 and dried at 60$^o$C in oven in ambient air.

The materials were annealed for 90 minutes in a tube furnace at 300$^o$C and 500$^o$C in ambient air (Lindberg/Blue M Mini-Mite™). The ramp rate was set at 10$^o$C.min$^{-1}$.

Samples were examined using a cold field-emission scanning electron microscope (SEM, JEOL JSM-7500F) and a high resolution transmission electron microscope (HRTEM, JEOL JEM-2200FS, operated at 200kV). X-ray energy-dispersive spectrometry (XEDS) in the SEM and the TEM was used to determine the elemental composition of the samples. Selected area electron diffraction (SAED) and powder x-ray diffraction (XRD) were employed to determine the phase(s) present. Electron energy-loss spectra (EELS) were recorded using the *in-column* energy filter fitted on the TEM. X-ray photoelectron spectroscopy (XPS) measurements were performed using a commercial PHI 5500 spectrometer with a monochromatized Al $K_\alpha$ radiation (1487eV) and an electron emission angle of 45°.

**Results and Discussion**

One SEM and two TEM images recorded from the TNPs ($WO_3 \cdot H_2O$) are displayed in Figure 1. They were less than 30 nm. XEDS recorded in the SEM and the TEM showed the materials to be composed of only elements O and W (data not shown here).

The diffraction pattern (DP) shown in inset I of Figure 1(*a*) was indexed to the orthorhombic *Pmnb* structure of tungstite with lattice parameters a=0.524, b=1.070 and c=0.512 nm, which was in good agreement with XRD card number JCPDS No. 43-0679 and the American Mineralogist Crystal Structure Database, AMCSD 0005199 [34]. The structure consists of distorted octahedral units of tungsten atoms coordinated with five oxygen atoms and a water molecule where the octahedra share four vertices in the equatorial plane forming the sheet structure (*see ESI, Figure S1b*). One HRTEM image recorded from the center of one TNP and its corresponding SAED pattern are shown in Figure 1(b, c). The measured *d*-spacing was 0.35 nm, corresponding to {111} of the orthorhombic crystal phase. The DP was indexed to orthorhombic tungstite, along the [110] zone axis.

Figure 2(a) shows a TEM image recorded from annealed TNPs at 300$^o$C for 90 min. The inset is a SAED pattern, indexed to monoclinic $WO_3$ demonstrating that the phase transformation to tungsten oxide occurred while preserving the nanoparticle morphology. Figure 2(b) shows one HRTEM image recorded from one $WO_3$ nanoparticle (WNP), where the (200) and (040) planes are labeled (0.36 and 0.19 nm, respectively). The SAED pattern shown in Figure 1(c) was recorded from one WNP and was indexed to monoclinic *P21/n (14)* $WO_3$, along the [001] zone axis, with lattice parameters a=0.729, b=0.754 and c=0.768 nm, which was in good agreement with published data ($WO_3$ x-ray diffraction card number JCPDS No. 43-1035). The same results were obtained with TNPs annealed at 500$^o$C (*See ESI, Figure S2, TNPs treated at 500$^o$C*). In this monoclinic structure, the W-O bonds make zigzag chains in three directions with W-O-W and O-W-O angles measuring 158° and 166°, respectively. In the *x* direction the bonds have equal lengths, while in the *y* and *z* directions they are alternately long and short [35]. XRD results were also in good agreement (*See ESI, Figure S1a*). Indeed, the phase transformation from tungstite to tungsten oxide is due to the removal of inter-structural water molecules (*See ESI, Figure S1b*). The formation of tungstite and its dehydration can be described in three separate steps [36]. The first two steps are schematically illustrated in *ESI, Figure S3*. The first step is the protonation of the tungstate ions upon solution acidification to form white solid precipitates:

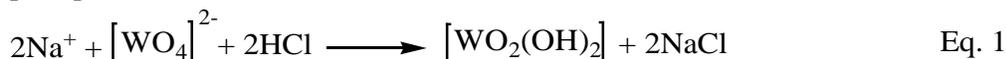

$$2Na^+ + [WO_4]^{2-} + 2HCl \longrightarrow [WO_2(OH)_2] + 2NaCl \qquad Eq. 1$$

Oxalic acid is a good surfactant to control the size of the final product [31] and makes the solution transparent. An amorphous gel of $WO_3.2H_2O$ (synthesized by mixing a $Na_2WO_4.2H_2O$ solution with 1N HCl solution) was reported to crystallize upon washing with water [37], however there was no control of the particle size. The second step is the hydration of the $[WO_2(OH)_2]$ tetrahedral molecules and dimerization via O-bridging to form crystalline $[WO(OH)_3(H_2O)]_2(\mu\text{-O})$ containing octahedral W-centers via the HT process in acidic solution:

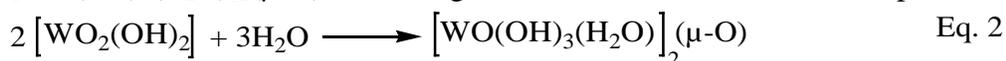

$$2[WO_2(OH)_2] + 3H_2O \longrightarrow [WO(OH)_3(H_2O)]_2(\mu\text{-O}) \qquad Eq. 2$$

An increase in the temperature during the nucleation process leads to a decrease in the activation energy, therefore promoting nucleation. However, Ostwald ripening results in a decrease in the final particle size. Therefore, during this step, the morphology and size of the tungstite materials can also be controlled using proper additives and chelating agents. For example, Gu *et al.* [28] reported on the synthesis of $WO_3$ nanorods in the presence of sulfate salts of alkali and alkaline earth metals using HT method at temperatures greater than 150°C inside an autoclave (high pressure). Shibuya *et al.* [29] synthesized tree like structures in the presence of rubidium sulfate at 150°C using the HT process. A high temperature HT process requires a stainless steel vessel lined with an acid resistive coating. Moreover, this process can take a long time (more than one day). In our study, we found that by adding sulfate ions during this step, nanorods were also produced (*See ESI, Figure S4*). The dominant growth direction in the presence of sulfate ions (here by adding 5mM sulfuric acid) is through {200} family planes. Also by adding 10mM $Cs^+$ (as $CsNO_3$) during this step, Cs-doped TNPs were obtained. Annealing preserved the Cs in the structure and Cs-doped $WO_3$ was achieved (*See ESI, Figure S5*). The Cs $M_{4,5}$ edge (EELS) is illustrated in *Figure S5*, indicating the presence of Cs atoms inside the structure of the TNPs. The presence of Cs in $WO_3$ lowers significantly its band-gap [38] and can be used as exchange by other elements to create proper active sites for water oxidation process [39].

The third step corresponds to the dehydration process (i.e., phase transformation):

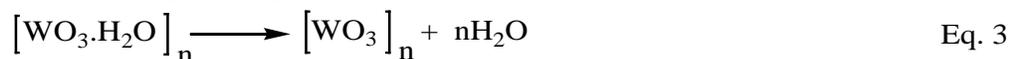

$$[WO_3.H_2O]_n \longrightarrow [WO_3]_n + nH_2O \qquad Eq. 3$$

The schematic of this dehydration process is shown in *ESI, Figure S1b*. The removal of internal water molecules and W=O double bonds allows for the layers to connect through the O atoms to stack together and turn to more compact structure.

The O1s and W4f core level XPS spectra recorded from TNPs and WNPs (heat treated TNPs at 300°C for 90 min.) samples are shown in Figure 3. The W*4f* spectra recorded from both samples present the W*4f*$_{7/2}$ and W*4f*$_{5/2}$ doublet at 35.6 and 37.7 eV (with peak separation of 2.1 eV and the intensity ratio of 3:4) associated with the oxidation state of $W^{6+}$ ions [40] (see Figure 3*a*). Bonding to oxygen atoms shifts these two peaks from the theoretical values of tungsten metal (31.4 and 33.6 eV). The small peak at 41.7 eV was assigned to W5p. The O*1s* spectrum of each sample contains a main contribution at the lowest binding energy, originated from W-O bond (see Figure 3*b*). The second peak, at about 532.5 eV, represents the –OH bond, which originates from the absorbed or inter-structural water molecules inside the tungsten oxide or the tungstite structure. The relative intensity of this peak is higher in the TNPs sample compared to that in the WNP sample because of the presence of inter-structural water molecules, removed during the dehydration process to form WNP (*see ESI, Figure S1b*). The relative intensity of this peak, 33%, is half of the relative intensity of the first peak (W-O peak) in the TNP spectrum.

EEL spectra were recorded near the W $N_{6,7}$, W $O_{1,2}$ and the O K edges. The spectra in the region of the W $N_{6,7}$ and W $O_{1,2}$ edges are shown in Figure 4(a). The peaks below W $N_{6,7}$ (shoulder of W$4f_{7/2}$) are primarily associated with collective excitations from the valence bands (VB) to the conduction bands (CB). According to the molecular orbital (MO) theory, the VB consists of four orbital components ($a_{1g}$, $t_{1u}$, $e_g$ and $t_{2g}$) to anti-bonding orbital components (*See ESI, Figure S6*).

In Figure 5, the EEL spectrum recorded from $WO_3$ in the low energy-loss region was fitted with Lorentzian peaks: two peaks at the energies of 5.0, 6.8 eV (Figure 5(a)) and four peaks at 12.6, 16.4, 21.8 and 26.2 eV (Figure 5(b)). The peak at 26.2 eV corresponds to carbon [41] (from the ultra-thin carbon supporting film of the copper TEM grid; about 3 to 5nm thick according to the vendor). Considering that the peaks below W $N_{6,7}$ are associated with possible excitations between the valence and conduction bands, and also the fact that exciting electrons from the *g* orbital components to $e_g^*$ to $t_{2g}^*$ anti-bonding components is forbidden, the first two low energy peaks are the result of the electron excitation from $t_{1u}$ occupied orbitals to two unoccupied anti-bonding components $e_g^*$ and $t_{2g}^*$ resulting to 5.0 and 6.8 eV peaks (*See ESI, Figure S6*). This is in agreement with UV/Vis studies [33] which show the peak separation for these two electron excitations to be 1.7 eV. On the other hand, mind that these two peaks may be affected by the zero-loss peak resulting in inaccurate measurements. The energies at 12.6, 16.4 and 21.8 eV can be assigned to the electron excitation from $t_{1u}$ to $a_{1g}^*$, $t_{2g}/e_g$ to $t_{1u}^*$ and $a_{1g}$ to $t_{1u}^*$, respectively (*See ESI, Figure S6*). For tungstite, these peaks cannot be assigned accurately due to the large carbon peak at 26.2 eV.

The O K-edge can be influenced by several factors such as the ejection of electrons into the continuum levels, making a saw-tooth shape, while solid state effects and transitions to unoccupied bonds create the fine structures in the edge [42-43]. The energy-loss near-edge structure (ELNES) at the O K-edge in transition metal oxides is due to the crystal-field splitting of the metallic *d* orbital into $e_g$ and $t_{2g}$ components, creating a double-peak signature [42-44] as shown in Figure 4(b). The peaks' positions and peak separation between the two first peaks ($\Delta_{1-2}$) for TNPs and WNPs (heat-treated at 300°C) were measured and are listed in Table I. The first peak's position (~1.20 eV) for both samples was not affected. However, the second peak in TNPs (6.25 eV) was significantly displaced compared to that of WNPs (3.80 eV). The third peak's energy for TNPs (8.00 eV) was less than that of WNPs (10.05 eV). Peaks at higher energy losses should be considered carefully because of the contributions of direct transitions to unoccupied states and multiple scattering resonances [44]. The first two peaks are due to the $e_g$ and $t_{2g}$ components and their separation ($\Delta_{1-2}$) is given in Table I. $\Delta_{1-2}$ was larger for TNPs.

McComb and Harvey reported that the oxygen ELNES of each phase had the same general shape, but that there exist significant differences in the locations and intensities of the individual peaks comprising the ELNES, and that the separation between two peaks was structure-dependent [44 and 43]. The monoclinic structure has a higher density and lower degree of symmetry than the orthorhombic structure, explaining the observed differences. McComb's analysis showed that the observed decreased peak separation is a result of the increased structural complexity, polyhedron variety, and reduced symmetry. These parameters may affect the initial O-$1s$ state and the splitting of the metal $5d$ level, which is hybridized with the final O -$2p$ state. Therefore, EELS (unlike XEDS) can be used to distinguish between tungstite and tungsten oxide. The band gap ($E_g$) for WNPs sintered at 300$^o$C was measured to ~2.6 eV according to McComb's extrapolation method.

**Conclusions**

Orthorhombic tungstite nanoparticles (less than 30 nm) were prepared using the acid precipitation method followed with a low temperature and low pressure hydrothermal treatment. Morphologies can be tuned by making additions during the HT treatment (e.g., the presence of sulfate ions resulted in rods). A short annealing to 300$^o$C in ambient air allowed for the phase transformation to monoclinic WO$_3$, while preserving the morphology. EELS can be used to distinguish between the two phases (unlike XEDS). This method of preparation is easily scalable for industrial applications and does not require specialty vessels.

**Acknowledgements**

The partial support from NSEC Center for Hierarchical Manufacturing at the University of Massassuchets (National Science Foundation award 1025020). NSF for its support (award 0701525) to the Nanoscopy Facility, an electron microscopy facility at UPR.

**Table/Figure captions:**

**Table I.** Peak energies (eV) measured from the O K-edge of TNPs and WNPs (heat treated at 300$^o$C).

**Figure 1.** (a) SEM image showing the TNPs where the insets *I* and *II* show one TEM image and the corresponding SAED pattern indexed to orthorhombic tungstite, respectively. (b) HRTEM image recorded from the center of a TNP. The {111} planes are labelled. (c) Corresponding SAED pattern, indexed to orthorhombic tungstite.

**Figure 2.** (a) TEM image recorded from TNPs annealed at 300$^o$C for 90 min. The inset is a SAED pattern, indexed to monoclinic WO$_3$. (b) HRTEM image recorded from one WO$_3$ nanoparticle. The lattice is monoclinic with the (200) and (040) planes labeled (0.36 and 0.19 nm, respectively). (c) SAED pattern recorded from one WNP, indexed to monoclinic WO$_3$ along the [001] zone axis.

**Figure 3.** O*1s* and W*4f* core level XPS spectra recorded from TNPs and WNPs (annealed at 300$^o$C). Intensities are normalized.

**Figure 4.** EELS recorded from a few TNPs and WNPs (annealed at 300$^o$C) at low energy-loss (a) and O k-edge (b) regions.

**Figure 5.** Experimental EELS and fitting for sample WNPs (annealed at 300$^o$C).

**Table I.**

| Sample | Peak Position (eV) | | | $\Delta_{1-2}$ (eV) |
|---|---|---|---|---|
| | 1 | 2 | 3 | |
| **WNPs - 300** | 1.15 | 3.80 | 10.05 | 2.65 |
| **TNPs** | 1.20 | 6.25 | 8.00 | 5.05 |

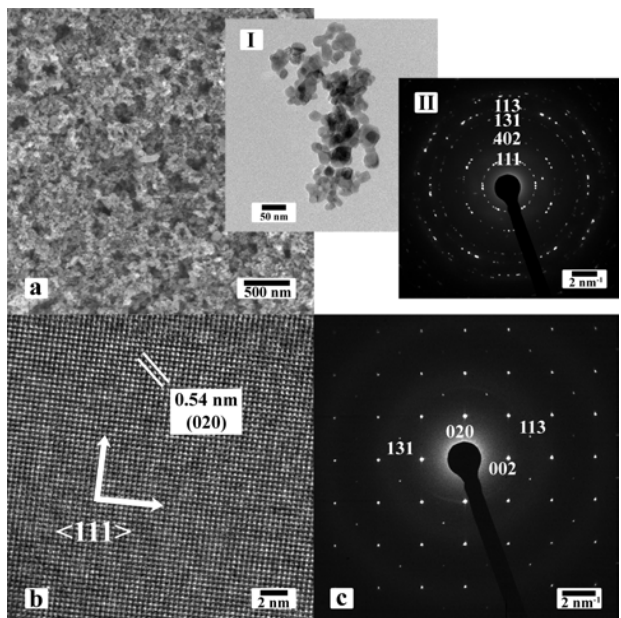

**Figure 1.**

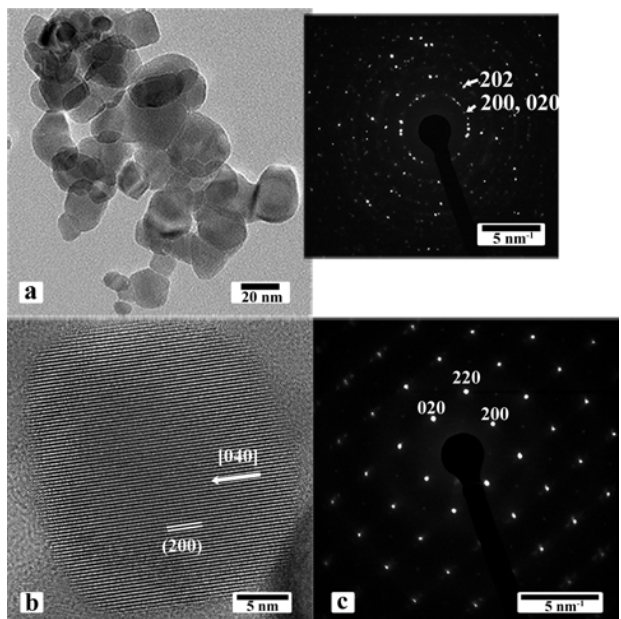

**Figure 2.**

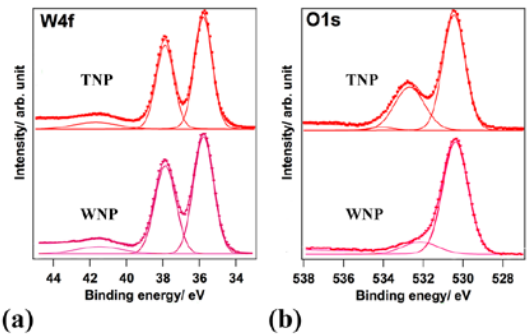

**Figure 3.**

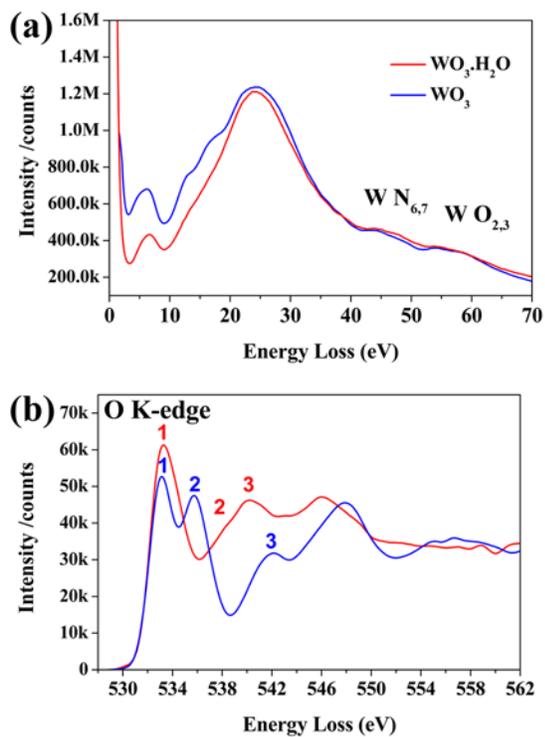

**Figure 4.**

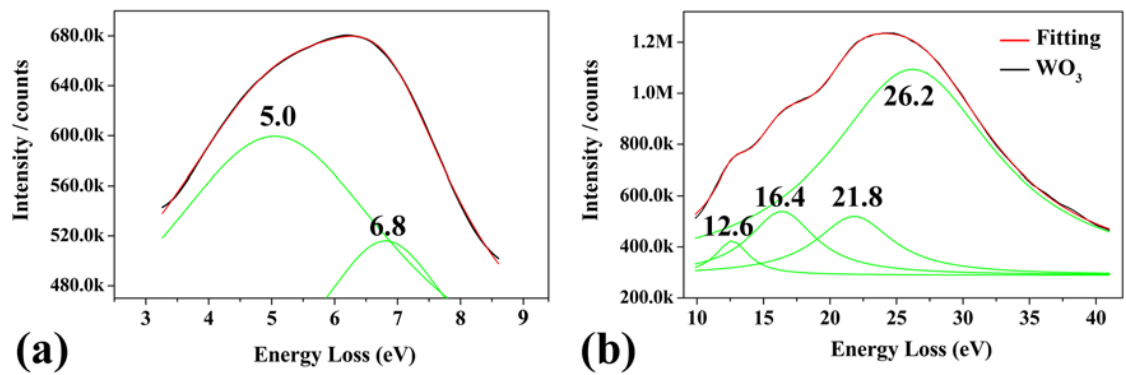

**Figure 5.**

*Synthesis of Tungsten Oxide Nanoparticles using a Hydrothermal Method at Ambient Pressure. M. Ahmadi, R. Younesi and M. Guinel*